\documentclass[aps,prd,twocolumn,floatfix,nofootinbib,showpacs,superscriptaddress,tightenlines]{revtex4}
\usepackage{graphicx}
\usepackage{color}
\usepackage{amsbsy}
\usepackage{color}
\usepackage{epsfig}
\usepackage{graphicx}
\usepackage{amsmath}
\usepackage{amssymb}
\usepackage{bbm}
\usepackage{amsbsy}
\usepackage{slashed}

\usepackage{xcolor}

\newcommand{\beq}{\begin{equation}}
\newcommand{\eeq}{\end{equation}}

\definecolor{mbcol}{rgb}{1,0,1}

\usepackage[normalem]{ulem}

\begin{document}
\title{Color, Flavor, Temperature and Magnetic Field Dependence of QCD Phase Diagram: Magnetic Catalysis and its Inverse}

\author{Aftab Ahmad}
\affiliation{
Institute of Physics and Electronics, Gomal University, 29220 D.I. Khan, K.P.K., Pakistan.}

\author{Adnan Bashir}
\affiliation{Instituto de F\'{\i}sica y Matem\'aticas, Universidad Michoacana de San Nicol\'as de Hidalgo,
Edificio C-3, Ciudad Universitaria, C.P. 58040, Morelia, Michoac\'an, M{\'e}xico.}

\author{Marco A. Bedolla}
\affiliation{
Mesoamerican Centre for Theoretical Physics, Universidad Aut\'onoma de Chiapas,
Carretera Zapata Km. 4, Real del Bosque (Ter\'an), Tuxtla Guti\'errez 29040, Chiapas, M\'exico.}

\author{J.~J. Cobos-Mart\'{\i}nez}
\affiliation{
C\'atedra CONACyT, Departamento de F\'{\i}sica, Centro de
Investigaci\'on y de Estudios Avanzados del Instituto Polit\'ecnico Nacional, Apartado Postal 14-740, 07000, Ciudad de M\'exico, M\'exico.}

\begin{abstract}
We study dynamical chiral symmetry breaking for quarks in the fundamental
representation of $SU(N_c)$ for $N_f$ number of light quark flavors.
We also investigate the phase diagram of quantum chromodynamics at finite temperature $T$
and/or in the presence of a constant external magnetic
field $eB$. The unified formalism for this analysis is provided by a
symmetry-preserving Schwinger-Dyson equations treatment of a
vector$\times$vector contact interaction model which encodes several well-established
features of quantum chromodynamics to mimic the latter as closely as possible. Deconfinement
and chiral symmetry restoration are triggered above a critical value of $N_f$ at $T=0=eB$.
On the other hand, increasing temperature
itself screens strong interactions, thus ensuring that a smaller value of $N_f$ is
sufficient to restore chiral symmetry at higher temperatures. We also observe the
well-known phenomenon of magnetic catalysis for a strong enough magnetic field. However, we note
that if the effective coupling strength of the model decreases as a function of
magnetic field, it can trigger inverse magnetic catalysis in a certain window
of this functional dependence. Our model allows for the simultaneous onset of dynamical chiral symmetry
breaking and confinement for each case. Qualitative as well as quantitative
predictions of our simple but effective model are in reasonably satisfactory agreement with lattice results
and other reliable and refined predictions based upon intricate continuum studies of quantum chromodynamics.
\end{abstract}


\maketitle

\section{\label{section-1} Introduction}
Quantum chromodynamics (QCD) is the theory of  strong color force among quarks
and gluons. Ultraviolet and infrared peculiarities of QCD hinge upon an
interplay between the number of light quark  flavors $N_f$ and the number of
colors $N_c$ of the gauge group $SU(N_c)$. Sign of the one-loop
$\beta$-function depends upon the factor $-11N_c + 2N_f $. In real life QCD,
this sign is negative as $N_c = 3$ outmuscles $N_f$ and hence the strong
interactions are asymptotically free~\cite{Gross:1973id,Politzer:1973fx,
Gross:1973ju}. Only in a universe with $N_f > (11/2)N_c$, this
effect will be reversed and asymptotic freedom will be lost. Infrared QCD is
even more eerie. It exhibits emergent phenomena of dynamical chiral symmetry
breaking (DCSB) and confinement which are inconceivable in any perturbative approach
to QCD.

Modern lattice analyses appear to demonstrate that the restoration of a
chirally symmetric phase takes place somewhere between
$N_f\approx 8-12$~\cite{Hayakawa:2010yn,Cheng:2013eu,Appelquist:2014zsa,
Hasenfratz:2016dou,Appelquist:2018yqe}. Continuum studies of QCD back these claims and confirm
that the phenomena of confinement and DCSB also owe themselves
to the intricate balance between
$N_f$ and $N_c$. For $N_c = 3$, if the number of flavors exceeds
$N_f \approx 7-12$, quarks are deconfined and chiral symmetry is restored,
see for example
Refs.~\cite{Appelquist:1996dq,Appelquist:1999hr,Bashir:2013zha,Hopfer:2014zna,Doff:2016jzk,Binosi:2016xxu}.

The infrared behavior of QCD is also affected in the presence of a heat bath.
At low temperatures, the observable degrees of freedom continue to be color-singlet hadrons, whereas at high
temperatures, the interaction gets increasingly screened and weak,
causing a hadron's constituents to deconfine into a phase where the dominant
degrees of freedom are the defining ingredients of perturbative QCD, namely,
quarks and gluons. Increasing $T$ also triggers the transition of quarks with
large effective constituent-like masses to quarks with only current masses. Needless
to say, a quantitative study of this behavior has  been widely carried out
in literature, see for example Refs.~\cite{Bernard:2004je,Cheng:2006qk,Aoki:2009sc,Borsanyi:2010bp,Bazavov:2011nk,Levkova:2012jd,deForcrand:2014tha, Bhattacharya:2014ara,Bazavov:2017xul}
for lattice QCD and Refs.~\cite{Qin:2010nq,Fischer:2011mz,Ayala:2011vs,Gutierrez:2013sta,Gao:2015kea,Eichmann:2015kfa,Gao:2016qkh,Fischer:2018sdj,Shi:2020uyb} for works based upon continuum techniques of Schwinger-Dyson equations (SDEs).

It is also well known that in the presence of strong magnetic fields, it is
possible to generate fermion masses for any value of the coupling strength. This
phenomenon was first studied in quantum electrodynamics (QED) and was dubbed as magnetic
catalysis, see for example~\cite{Gusynin:1994xp,
Lee:1997zj,Hong:1997uw,Ferrer:2000ed,Ayala:2006sv,Rojas:2008sg,Ayala:2010fm}. This phenomenon
owes itself to dimensional reduction.
Nonperturbative aspects of dynamical mass generation in the presence of magnetic
fields have also been studied in continuum QCD~\cite{Kabat:2002er,
Miransky:2002rp}, as well as lattice QCD~\cite{DElia:2010abb}.

QCD phase diagram has also been studied in the presence of external magnetic fields.
It is observed that near the cross-over temperature, chiral quark condensate develops
a peculiar behaviour. It starts decreasing with increasing magnetic field.
This effect has come to be known as inverse magnetic
catalysis (IMC),~\cite{Bali:2011qj,Bali:2012zg,Bali:2013esa,Bornyakov:2013eya,Pagura:2016pwr},
or magnetic inhibition,~\cite{Fukushima:2012kc}. Continuum QCD studies support
these findings,~\cite{Mueller:2015fka}. It is important to highlight that
this effect is believed to be triggered both by the weakening of the
running coupling and the gluon dynamics. Therefore, any model-building must
incorporate important QCD features to stand any reasonable chance to capture
the correct behavior of this theory at finite temperature in the presence of external
magnetic fields. This is precisely what our contact interaction (CI) model does.
It is designed to meet the following requirements:
 \begin{enumerate}
     \item It produces right amount of DCSB at $T=0=eB$. Consequently,
     it has extensively been applied to satisfactorily calculate static properties
     of light and heavy
     hadrons,~\cite{GutierrezGuerrero:2010md,Roberts:2010rn,Chen:2012qr,Roberts:2011wy,Roberts:2011cf,Bedolla:2015mpa,Bedolla:2016yxq,Raya:2017ggu,Gutierrez-Guerrero:2019uwa}.
     \item It takes into account an infrared mass scale connected to the gluon mass
     which is known to emerge in non perturbative QCD,~\cite{Bowman:2004jm,Dudal:2008sp,Huber:2010cq,Boucaud:2011ug,Ayala:2012pb}.
     \item The color and light flavor dependence of DCSB and confinement are suitably
     built-in to mimic the recent QCD findings both on the lattice and in continuum
     QCD studies,~\cite{Hayakawa:2010yn,Cheng:2013eu,Appelquist:2014zsa,
Hasenfratz:2016dou,Appelquist:2018yqe,Appelquist:1996dq,Appelquist:1999hr,Bashir:2013zha,Hopfer:2014zna,Doff:2016jzk,Binosi:2016xxu}.
     \item It implements confinement by ensuring the absence of quark production thresholds.
     \item It supports the phenomenon of magnetic catalysis at zero temperature and inverse magnetic
     catalysis at finite temperature~\cite{Ahmad:2016iez}. We improve and generalize this model to study the dependence of
     these phenomena on the massless quark flavors. We require both the emergent phenomena
     of DCSB and confinement to be interlinked and simultaneous as evidence
     suggests~\cite{Bornyakov:2013eya,Mueller:2015fka}.
 \end{enumerate}

Equipped with a carefully constructed QCD based model under extreme conditions, we investigate its phase
diagram as a function
of colors $N_c$, light quark flavors $N_f$, finite temperature $T$ and external magnetic field $eB$.
All the predictions align satisfactorily with modern findings of the QCD phase diagram while avoiding
computational complexities of lattice and refined continuum studies.

In Sec.~II, we recall the quark gap equation, define the CI model and
solve it as a function of $N_c$ and $N_f$. For each $N_c$, there is a  critical value of $N_f$ above
which chiral symmetry is restored. After the screening dynamics has been adequately incorporated into the
CI, we study the QCD phase diagram in Sec.~III and obtain the cross-over temperature for chiral
symmetry restoration and deconfinement. Sec.~IV is devoted to the same study with an external
magnetic field replacing the thermal bath. We are able to describe magnetic catalysis with
its known characteristics. In Sec.~V, DCSB and confinement are studied as a function of
external magnetic field, temperature and light quark flavors. We find IMC in a region
including the cross-over temperature. Sec.~VI provides conclusions and discussion.

\section{\label{section-2} The Gap Equation}
We start by presenting the generalities of the CI for the quark fields.
In order to include the anti-screening effects of the gluons and the
screening effects of the light quarks, we extend the model to an
$SU(N_c)$ gauge theory with $N_f$ number of light quarks.

\subsection{The Contact Interaction}\label{sec:CImodel}
The dressed-quark propagator $S$ is obtained by solving the quark SDE
\begin{eqnarray}
S^{-1}(p)&=&i\gamma \cdot p + m_f \nonumber\\
&&\hspace{-1.5cm}
+ \int \frac{d^4q}{(2 \pi)^4} \; g^2 D_{\mu\nu}(p-q)\frac{\lambda^a}{2}\gamma_\mu S(q)
\Gamma_\nu^a(p,q)\,,
\label{eqn:gap-QCD}
\end{eqnarray}
where $m_f$ is the bare quark mass, $g$ is the QCD coupling constant, $\lambda^a$ are the Gell-Mann
matrices, $\Gamma_\nu^a $ is the dressed quark-gluon vertex and
$D_{\mu\nu}$ is the gluon propagator.

As we have mentioned before, in a series of
previous articles, it has been shown that at zero temperature, the static
properties of low-lying mesons and baryons can be faithfully reproduced by
assuming that the quarks interact, not via massless vector-boson exchange, but
instead through a symmetry preserving vector-vector CI with
a finite gluon mass~\cite{Chen:2012qr,Roberts:2011wy,Roberts:2011cf,Roberts:2010rn,GutierrezGuerrero:2010md,Bedolla:2015mpa,Bedolla:2016yxq,Raya:2017ggu}:
 \begin{eqnarray}
  g^2D_{\mu\nu}(k) &=& \delta_{\mu\nu} \frac{4\pi
  \alpha_{\rm IR}}{m_G^2} \equiv \delta_{\mu\nu}
  \alpha_{\rm eff}\,,\label{eqn:CImodel} \\
 \Gamma_\mu ^a (p,q)&=& \frac{\lambda^a}{2}\gamma_\mu\,\label{eqn:RLapp},
 \end{eqnarray}
\noindent where $m_G=500$ MeV is an infrared gluon mass scale which is generated
dynamically in QCD~\cite{Bowman:2004jm,Dudal:2008sp,Huber:2010cq,Boucaud:2011ug,Ayala:2012pb,Gao:2017uox}, and $\alpha_{\rm IR}=0.36 \pi$
specifies the strength of the infrared interaction. There is a
critical value of $\alpha_{\rm eff}$ above which chiral symmetry is dynamically broken.

 Eqs.~(\ref{eqn:CImodel}) and (\ref{eqn:RLapp}) specify the kernel in
 the quark SDE, Eq.~(\ref{eqn:gap-QCD}). In this approximation, the
 dressed-quark propagator takes a very simple form~:
\begin{equation}
S^{-1}(p)=i\gamma\cdot p+M,\label{eqn:gapCI}
\end{equation}
\noindent where $M$ is momentum independent dynamical quark mass, to be
determined from Eq.~(\ref{eqn:gap-QCD}). If we substitute Eqs.~(\ref{eqn:CImodel},\ref{eqn:RLapp},\ref{eqn:gapCI})
into Eq.~(\ref{eqn:gap-QCD}), and recall that in the fundamental representation of $SU(N_c)$, the Gell-Mann matrices
satisfy the identity
$\sum^{8}_{a=1} \lambda^{a} \lambda^{a}=2
\left(N_c-{1}/{N_c}\right)$, we get
\begin{equation}
M=m_f+\frac{ \alpha_{\rm eff}^{N_c}(N_f) M }{8\pi^2}\int^{\infty }_{0}
ds\frac{s}{s+M^2} \,, \label{eqn:PTRa}
\end{equation}
where
\begin{equation}
\alpha_{\rm eff}^{N_c}(N_f)= \left(N_c-{1}/{N_c}\right) \, \alpha_{\rm eff}(N_f) \,.
\label{eqn:alpha-flavor}
\end{equation}
Since the integral in Eq.~(\ref{eqn:PTRa}) is divergent, we must adopt a
regularization procedure. After exponentiation of the denominator of the
integrand and employing the confining proper-time regularization~\cite{Ebert:1996vx}, we can write
\begin{eqnarray}
\frac{1}{s+M^2}&=&\int^{\infty }_{0} d\tau {\rm e}^{-\tau(s+M^2)}
\rightarrow \nonumber\\
\int^{\tau_{\text{IR}}^{2}}_{\tau_{\text{UV}}^{2}} d\tau {\rm
e}^{-\tau(s+M^2)}
&=&\frac{ {\rm
e}^{-\tau_{\text{UV}}^{2}(s+M^2)}-{\rm e}^{-\tau_{\text{IR}}^{2}(s+M^2)}}{s+M^2}
. \label{eqn:PTRb}
\end{eqnarray}
\noindent Here, $\tau_{\text{IR,UV}}^{-1}=\Lambda_{\text{IR,UV}}$ are
infra-red and ultra-violet regulators, respectively. A non-zero value for
$\tau_{\text{IR}}$ implements confinement by ensuring the absence of quarks
production thresholds ~\cite{Roberts:2007ji}.  It has been shown that an
excitation described by a pole-less propagator would never reach its
mass-shell~\cite{Ebert:1996vx}. Moreover, since
Eq.~(\ref{eqn:CImodel}) does not define a renormalizable theory,
$\Lambda_{\text{UV}}$ cannot be removed, but instead plays a dynamical role,
setting the scale for all dimensioned quantities. After integration over $s$,
the gap equation reads~:
\begin{equation}
  \label{eqn:const_mass_reg}
 M = m_f + \frac{\alpha_{\rm eff}^{N_c}(N_f)}{8\pi^{2}}M
  \int^{\tau_{\text{IR}}^2}_{\tau_{\text{UV}}^2} d\tau \tau^{-2}
  {\rm e}^{-\tau M^2}\,.
\end{equation}
We shall use the notation $\alpha_{\rm eff}^{N_c}(N_f)=\alpha_{\rm eff}(N_f)$ for
$N_c=3$. Moreover, $\alpha_{\rm eff}(N_f)=\alpha_{\rm eff}$ for $N_f=2$.
We employ the parameters of Ref.~\cite{Chen:2012qr}, namely, we fix the coupling
to:
\begin{equation}
\alpha_{\rm eff}= 5.7\times 10^{-5}~\mathrm{MeV^{-2}},\label{p1}
\end{equation}
\noindent and use the following infrared and ultraviolet cutoffs:
\begin{equation}
       \tau_{\text{IR}} = (240~\mathrm{MeV})^{-1}, \qquad
       \tau_{\text{UV}} = (905~\mathrm{MeV})^{-1}.
       \label{p2}
\end{equation}
These parameters have been fitted to the value of the chiral quark condensate in the
vacuum. Along with $m_{f}=0$, we obtain $M=358$~MeV and $\langle\bar{u}u\rangle^{1/3} = \langle\bar{d}d\rangle^{1/3} = -241$~MeV
for the dynamical mass of the $u/d$ quarks and the chiral quark condensate
for two quark flavors  ($N_{f}=2$, $N_c=3$), respectively.

\subsection{Colors, flavors and chiral symmetry breaking}\label{sec:CriticalN}
%
%
\begin{figure}[t!]
\begin{center}
\includegraphics[scale=0.33]{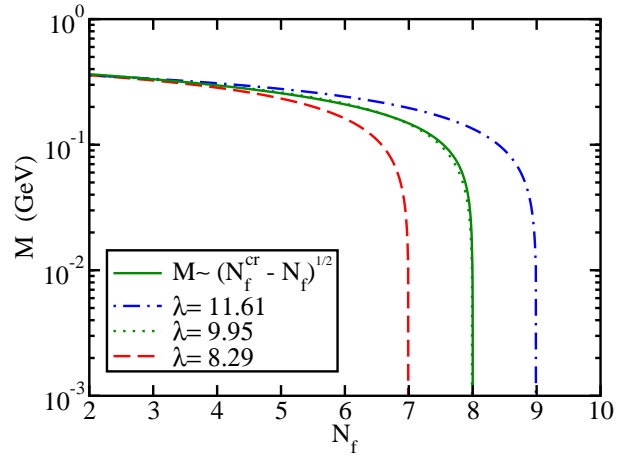}
\caption{Dynamical quark mass in the chiral limit as a function of flavors
$N_f$ with $N_c=3$ for three different values of $\lambda$ which ensures dynamical mass vanishes at $N_{f}^{\text{cr}}=7,8$ and $9$.}
\label{Fig1}
\end{center}
\end{figure}
%
%
\begin{figure}[t!]
\begin{center}
\includegraphics[scale=0.33]{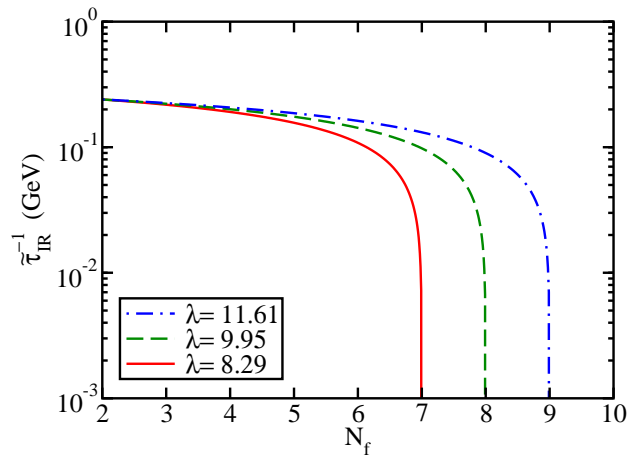}
\caption{Confining length scale in the chiral limit as a function
of $N_f$ flavors with $N_c=3$ colors. The  critical number of flavors
for  confinement are $N_{f}^{\text{cr}}=7, 8$ and $9$.}
\label{Fig2}
\end{center}
\end{figure}
 To study the effect of light quark flavors $N_f$ on the DCSB, we adopt the
 following flavor-dependence of the
 effective coupling:
\begin{equation}
\alpha_{\rm eff}^{N_c}(N_{f})=\alpha_{\rm eff}^{N_c} \sqrt{1-\frac{(N_{f}-2)}{\lambda}}.
\label{fc}
\end{equation}
This form owes itself to the pattern of DCSB as observed
in~\cite{Bashir:2013zha}. The dynamical quark mass in that work was observed to
be $m_{\rm dyn} \sim \sqrt{1 - N_f/ N_f^c}$ where $N_f^c$ is the critical number of
flavors above which chiral symmetry is restored (see Eq.~(8) of that article).
Only a similar square root dependence in the coupling, Eq.~(\ref{fc}), leads to this observed behavior.
Thus the functional form in Eq.~(\ref{fc}) can be traced back to the flavor dependence of the gluon
propagator in QCD,~\cite{Ayala:2012pb}. Due to this direct connection, note that $\lambda \sim N_f^c + \delta$,
where $\delta \sim 1.2-2.6$ for the cases of interest to us.
The appearance of $\delta$ is due to the factor $N_f-2$ in Eq.~(\ref{fc}).

Employing this coupling, we solve the gap equation,
Eq.~(\ref{eqn:const_mass_reg}).  For $N_c=3$, we identify different values of
$\lambda$ to obtain critical values of
$N_f$ ($N_{f}^{\text{cr}}=7, 8, 9$) above which the screening effects restore
chiral symmetry. These numbers are in accordance with the findings of Ref.~\cite{Bashir:2013zha}, later
reproduced in~\cite{Binosi:2016xxu}.

In Fig.~\ref{Fig1}, we display the variation of the dynamical mass $M$ as a
function of the number of flavors $N_f$. We compare one of the plots
with a fit given by $M \sim \sqrt{N_f^{\text{cr}}-N_f}$,
see~\cite{Bashir:2013zha}, for $N_f^{cr}= 8$. It corresponds to
$\lambda= 9.95$.
The numerical results are very well described by this analytic form,
reaffirming that our CI model mimics the refined continuum studies
presented in~\cite{Bashir:2013zha}. For the remainder of the article, we
choose the parameters to reproduce $N_f^{cr}= 8$ unless otherwise mentioned.

Recall that the confinement parameter is $\tau_{\text{IR}}$. Simultaneous DCSB
and confinement can be readily incorporated into the model if we define a
flavor dependent infrared cutoff~:
\begin{equation}
\widetilde\tau_{\text{IR}}=\tau_{\text{IR}} \frac{M(2)}{M(N_{f})}\,, \label{eqn:gapf}
\end{equation}
where $M(2)$ is the dynamical mass when $N_{f}=2$ and $N_{c}=3$.
In Eq.~(\ref{eqn:gapf}), $M(N_f)\rightarrow 0$, implies
$\widetilde\tau_{\text{IR}}\rightarrow \infty$ and hence quarks get deconfined
in  this model. Seen in conjunction with Fig.~\ref{Fig1}, plots in
Fig.~\ref{Fig2} provide a confirmatory numerical check that DCSB and
confinement are simultaneous in our CI model.

\begin{figure}[t!]
\begin{center}
\includegraphics[scale=0.32]{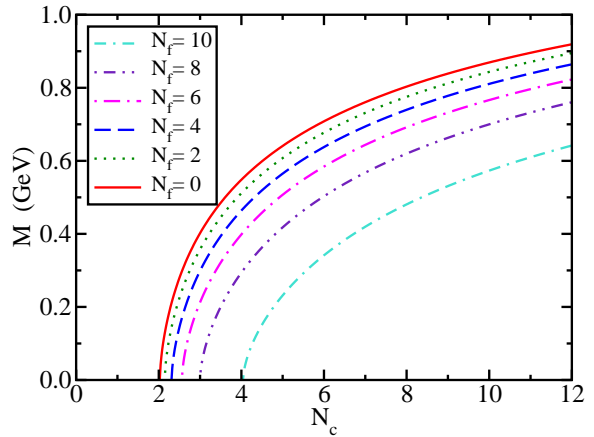}
\caption{Dynamical mass in the chiral limit as a function of $N_c$ for different values of $N_f$.
With increasing $N_f$, higher $N_c$ is required to trigger DCSB. For $N_f=0$,
the dynamical mass vanishes below
$N_{c}^{\text{cr}}\approx 2$.}
\label{Fig3}
\end{center}
\end{figure}

\begin{figure}[t!]
\begin{center}
\includegraphics[scale=0.32]{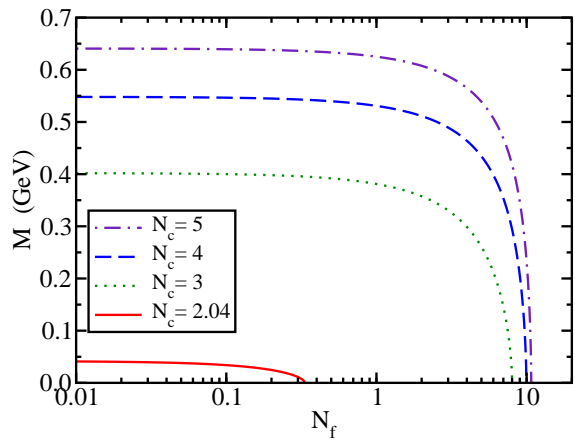}
\caption{Dynamical mass in the chiral limit as a function of quark flavors
for several values of $N_c$. For a given $N_c$, one can increase $N_f$ to a critical value
such that
the interaction strength diminishes enough to suppress DCSB.}
\label{Fig4}
\end{center}
\end{figure}

\begin{figure}[t!]
\begin{center}
\includegraphics[scale=0.32]{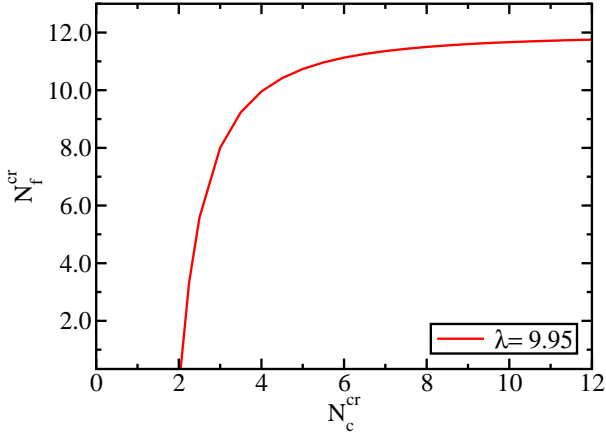}
\caption{Critical number of flavors $N_f^{\text{cr}}$ as a function of critical number of colors $N_c^{\text{cr}}$.
The above figure demonstrates the diametrically opposed effects of these two parameters.}
\label{Fig5}
\end{center}
\end{figure}

As emphasized earlier, the number of quark colors $N_c$ anti-screens the
interactions while $N_f$ screens them. This can be readily confirmed by
plotting the dressed quark mass $M$ for different number of flavors, see
Fig.~\ref{Fig3}. For $N_f=0$ the minimum value
of $N_c$ required to trigger DCSB is $N_c^{cr} \simeq 2$. As the number of
massless quark flavors increases, mellowing down the interaction strength,
higher color group $SU(N_c)$ has to be invoked to set off DCSB.
This interplay, reminiscent of the asymptotic freedom, is depicted in Figs.~\ref{Fig3}, \ref{Fig4}, \ref{Fig5}.
%
%
%
%
%
%

\section{DCSB at Finite Temperature $T$}
%
%
\begin{figure}[t!]
\begin{center}
\includegraphics[width=0.48\textwidth]{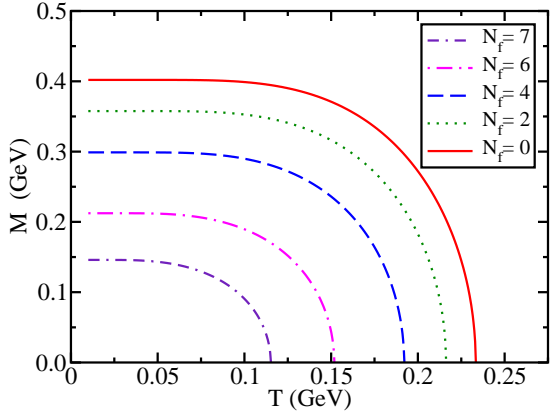}
\caption{Dynamical mass as a  function of $T$ for different number of flavors $N_{f}$. For $N_{f}\ge 8$
there is no generated mass.}
\label{Fig6}
\end{center}
\end{figure}
%
%
\begin{figure}[th!]
\begin{center}
\includegraphics[width=0.48\textwidth]{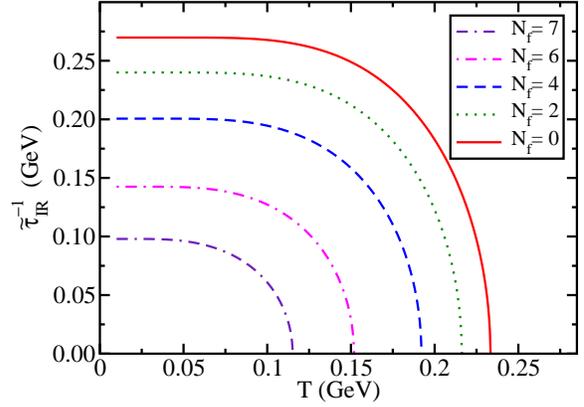}
\caption{Confining length scale as a function of $T$ for different values
of $N_f$.  Its behavior is very similar to the DCSB. Increasing temperature of the heat bath eventually restores
chiral symmetry.}
\label{Fig7}
\end{center}
\end{figure}
%
%
\begin{figure}[th!]
\begin{center}
\includegraphics[width=0.48\textwidth]{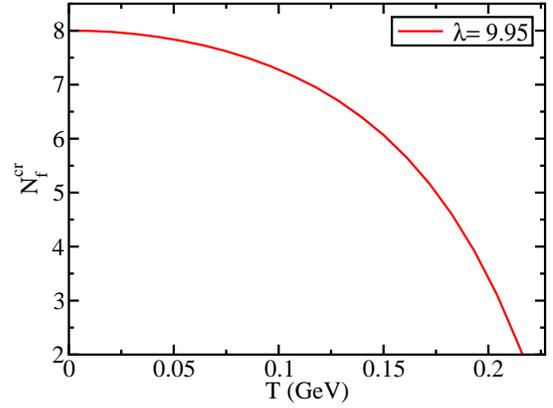}
\caption{The phase diagram  of chiral symmetry and confinement for
$N_{f}^{cr}$ vs $T=T_{c}$. The  $N_{f}^{cr}$ and $T_{c}$ are obtained from the
thermal gradient of the dynamical mass $\partial_T M$ and of the confining
scale, i.e., $\partial_T \widetilde{\tau}_{\text{IR}}^{-1}$. }
\label{Fig8}
\end{center}
\end{figure}
At finite temperature, within the imaginary time formalism, we split the
fermion four-momentum according to $q=(\omega_{n},\vec{q})$, where
$\omega_n = (2n+1)\pi T$ are the well-known fermionic Matsubara frequencies.
We adopt the standard convention for momentum integration, namely:
\begin{equation}
\int\frac{d^4q}{(2\pi)^4} \rightarrow T \sum^{\infty}_{n=-\infty}
 \int\frac{d^3q}{(2\pi)^3}. \label{mf}
\end{equation}
\noindent Thus, the gap equation in the chiral limit at finite temperature can
be written as (for $N_{c}=3$):
\begin{eqnarray}
 \hspace{-.4cm} M= \frac{8 \alpha_{\rm eff}(N_f) MT}{3\pi^2} \sum^{\infty}_{n=-\infty}
\int_0^\infty dq  \frac{\vec{q}^2}{\vec{q}^2 +\omega^{2}_{n} + M^2}\,.
\label{gapT1}
\end{eqnarray}
 This equation and some of its variants have been discussed in several
 works~\cite{Ahmad:2016iez,Marquez:2015bca,Cui:2014hya,Wang:2013wk}. To
 implement proper time regularization, we exponentiate the denominator for each
 $\omega_n$ as follows~:
\begin{equation}
\frac{1}{\vec{q}^2 +\omega^{2}_{n}+ M^2}\longrightarrow
\int^{\widetilde{\tau}^2_{\text{IR}}}_{\tau^2_{\text{UV}}}d\tau
{\rm e}^{-\tau(\vec{q}^2 +\omega^{2}_{n}+ M^2)},
\label{gapT2}
\end{equation}
with
\begin{equation}
\widetilde{\tau}_{\text{IR}}=\tau_{\text{IR}} \frac{M(0,2)}{M(T,N_{f})} \label{gapT3}\,,
\end{equation}
 where $M(0,2)$ is the dynamical mass at $T=0$, $N_f=2$ and $N_{c}=3$.
 Thus, in the chiral limit, the confining scale vanishes at the chiral symmetry
 restoration temperature. This is a simple way of ensuring the coincidence
 of transitions to confinement and DCSB phase. Summation over
 Matusbara frequencies and the remaining radial integration are carried out by using
 the following identities:
 \begin{eqnarray}
\sum^{\infty}_{n=-\infty} e^{-\tau\omega_{n}^2}&=&\Theta_{2}(0,e^{-(2\pi T)^{2}\tau})\;, \label{gapT4v1}\\
\int^{\infty}_{0} dq \; \vec{q}^{\;2} e^{-\tau\vec{q}^{\;2}}
&=& \frac{\sqrt{\pi}}{4 \tau^{3/2}},\label{gapT4v2}
\end{eqnarray}
where $\Theta_{2}(x,y)$ represents the second Jacobi theta function. Finally,
we arrive at the expression for the gap equation at finite
temperature~:
\begin{eqnarray}
\hspace{-0.5cm} M = \frac{2\alpha_{\rm eff}(N_f)MT}{3\pi^{3/2}}
\int^{\widetilde{\tau}^2_{\text{IR}}}_{\tau^2_{\text{UV}}}d\tau
\frac{{\rm e}^{-M^2\tau}\Theta_{2}(0, {\rm e}^{-(2\pi T)^{2}\tau})}{\tau^{3/2}}
\,. \label{gapT5}
\end{eqnarray}
\noindent
In this chiral limit, the thermal evolution of dynamical mass $M$, for
different values of $N_{f}$, is shown in Fig.~\ref{Fig6}. As expected, with
increasing temperature, the strong interaction gets weakened, and, therefore, a
lower value for the critical number of massless quark flavors is needed
to restore chiral symmetry for a given value of $N_c$ (3 in this case).
A similar behavior is expected and observed for the confinement scale
$\widetilde{\tau}_{\text{IR}}^{-1}$, see Fig.~\ref{Fig7}. Unlike the competing
forces of $N_c$ and $N_f$, temperature $T$ and the number of flavors $N_f$
both catalyze chiral symmetry restoration and deconfinement. Note that the
temperatures for the chiral symmetry breaking-restoration $(T_c^\chi)$,
confinement-deconfinement transitions ($T_{c}^{c}$) as a function of $N_{f}^{cr}$, shown in
Fig.~\ref{Fig8}, are determined, respectively, from the position  of the
divergences of their thermal gradients $\partial_T M$ and
$\partial_T\widetilde{\tau}_{\text{IR}}^{-1}$.
The model construction ensures these critical temperatures are coincidental,
$T_c^\chi=T_c^c\equiv T_c\simeq 216.5$~MeV for
$N_{f}=2$ and $N_{c}=3$~\cite{Ahmad:2016iez,Marquez:2015bca,Wang:2013wk}.

In the following section, instead of a thermal bath, we study the effect of a
constant and uniform external magnetic field on the DCSB and confinement within
this framework of the CI.

\section{Gap equation in a Magnetic field with $N_{c}=3$}
%
%
\begin{figure}[t!]
\begin{center}
\includegraphics[width=0.48\textwidth]{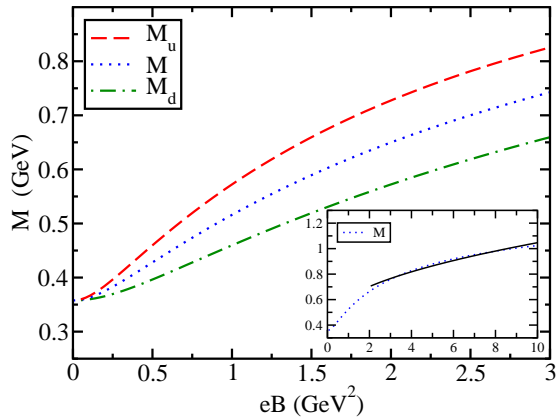}
\caption{ 
Dynamical mass in the chiral limit for $u$ (${\cal{Q}}_u=+2e/3$),
$d$ (${\cal{Q}}_d=-e/3$) quarks as a function of the strength
of the magnetic field, for $N_f=2$ and $N_{c}=3$. $M$ denotes the average
dynamical mass, Eq.~(\ref{eqn:mave}), obtained by solving its gap equation,
 Eq.~(\ref{eqn:mave})}
\label{Fig9}
\end{center}
\end{figure}

%
%
\begin{figure}[t!]
\begin{center}
\includegraphics[width=0.48\textwidth]{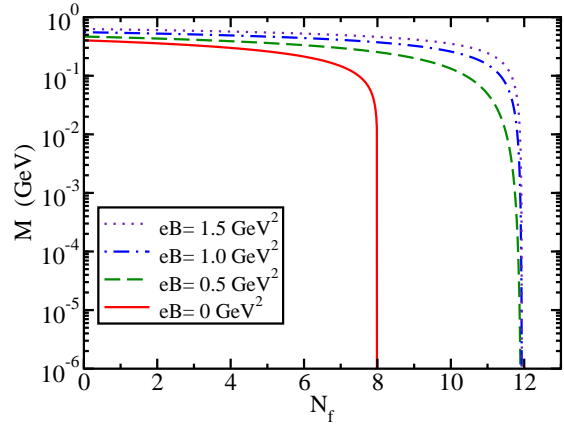}
\caption{Dynamical average mass $M$ as a function of $N_{f}$ and magnetic field strength $eB$ in the chiral limit for $N_{c}=3$.}
\label{Fig10}
\end{center}
\end{figure}
%
%
\begin{figure}[t!]
\begin{center}
\includegraphics[width=0.48\textwidth]{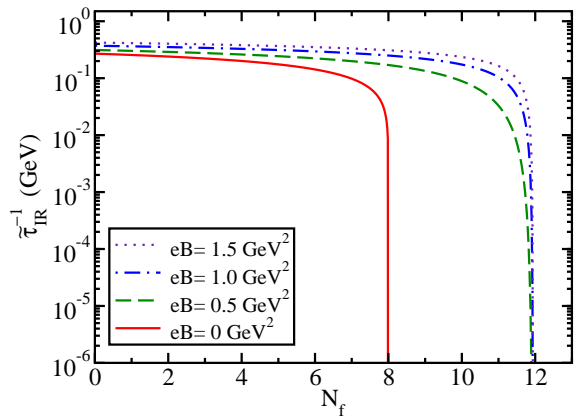}
\caption{Confining length scale as a function of $N_{f}$ and magnetic field strength $eB$ in the chiral limit for $N_{c}=3$.}
\label{Fig11}
\end{center}
\end{figure}
In this section we consider a background homogeneous magnetic field
directed along the $z$-axis, with magnitude $B$ and defined through the
symmetric gauge vector potential:
\begin{equation}
A^{ext}_{\mu}=\left( 0,-\frac{B y}{2}, \frac{B x}{2},0\right)\,.
\end{equation}
The quark propagator $S(q)$ gets dressed with magnetic field effects,
$S(q)\to \widetilde{S}(q)$, in the Fock-Schwinger
representation~\cite{Schwinger:1951nm,Fock:1937dy}, and it is:
\begin{eqnarray}
&&\hspace{-15mm}\widetilde{S}(q)=\int^{\infty}_{0} ds
\frac{{\rm e}^{-s(q^{2}_\parallel+q^{2}_{\perp}\frac{{\rm tan}(|{\cal{Q}}_{l} Bs|)}{|{\cal{Q}}_{l} Bs|}+M^{2})}}{{{\rm cosh}(|{\cal{Q}}_{l} Bs|)}}\nonumber \\
&\times& \bigg[\bigg({\rm cosh}(|{\cal{Q}}_{l}Bs|)-i\gamma^{1}\gamma^{2}{\rm sinh}(|{\cal{Q}}_{l}Bs|)\bigg)\nonumber\\
&\times&
(M-\slashed{q}_{\parallel})-\frac{\slashed{q}_{\perp}}{{\rm cosh}(|{\cal{Q}}_{l}Bs|)} \bigg], \label{sh}
\end{eqnarray}
where the parallel and transverse splitting of quark momenta is in reference to
the magnetic field direction, as usual\footnote{Recall that
$q^2=q_\parallel^2+q_\perp^2$, with $q_\parallel^2= q^{2}_0+q^{2}_{3}$ and
$q_\perp^2= q^{2}_1+q^{2}_{2}$.}, and ${\cal{Q}}_{l}\, ({\cal{Q}}_u=+2e/3$,
${\cal{Q}}_d=-e/3)$ refers to the electric charge of the light quarks, with
$l=u,\,d$.
With these  ingredients, we adopt the regularization procedure from the
previous section. The corresponding gap equation for the  dynamical mass at
zero temperature under the influence of a uniform magnetic field for the light quarks becomes:
\begin{eqnarray}
M_{l}&=& \frac{16 \alpha_{\rm eff}(N_{f})}{3} M_{l}
\int \frac{d^2q_{\perp}}{(2\pi)^2}\frac{d^2q_{\parallel}}{(2\pi)^2}
\nonumber\\
&& \times  \int_{\tau^{2}_{\text{UV}}}^{\widetilde{\tau}^{2}_{\text{IR}}}
d\tau {\rm e}^{-\tau(q^{2}_{\parallel}+q^{2}_{\perp}
\frac{{\rm tanh}(|{\cal{Q}}_{l}B\tau|)}
{|{\cal{Q}}_{l}B\tau|}+M_{l}^{2})}.
\label{Mag7}
\end{eqnarray}
On using the relations:
\begin{eqnarray}
 \int \frac{d^2 q_{\parallel}}{ (2\pi)^2}e^{-\tau q^{2}_{\parallel}} &=&
 \frac{1}{4\pi \tau},\nonumber \\
 \int \frac{d^2 q_{\perp}}{ (2\pi)^2} e^{-\tau q^{2}_{\perp}
\frac{{\rm tanh}(|{\cal{Q}}_{l}B\tau|)}{|{\cal{Q}}_{l}B\tau|}}
&=&\frac{|{\cal{Q}}_{l}B|}{4\pi{{\rm tanh}(|{\cal{Q}}_{f}B\tau|)}},
\end{eqnarray}
we obtain the gap equation for massless quarks at zero temperature in an external magnetic field
\begin{eqnarray}
M_{l} &=& \frac{ \alpha_{\rm eff}(N_{f})}{3\pi^{2}}|{\cal{Q}}_{l}B|
\int_{{\tau}^2_{\text{UV}}}^{\widetilde\tau^2_{\text{IR}}}d\tau
\frac{{M_{l}\rm e}^{-M_{l}^2\tau}}{\tau {\rm \tanh}(|{\cal{Q}}_{l}B|\tau)}\,, \label{EMag2}
\end{eqnarray}
where $l=u,d$. In Fig.~\ref{Fig9}, we show numerical results
for the dynamical mass of the $u$ and $d$ quarks in the chiral limit
as a function of $eB$. The so-called magnetic catalysis effect is
clearly seen in this figure, and its influence is bigger in the
case of the $u$ quark due to its larger charge, compared to the $d$
quark.

For our purposes, and in order to make the analysis flavor-independent, in the following we will work with the average dynamical mass in the chiral limit,
\begin{equation}
\label{eqn:mave}
M=\frac{1}{2}(M_{u}+M_{d})\,,
\end{equation}
whose gap equation is given by
\begin{equation}
\label{eqn:mavegap}
M=\frac{ \alpha_{\rm eff}(N_{f})}{3\pi^{2}}\frac{1}{2}
\sum_{l=u,d}
|{\cal{Q}}_{l}B|\int_{{\tau}^2_{\text{UV}}}^{\widetilde\tau^2_{\text{IR}}}
d\tau \frac{{M\rm e}^{-M^2\tau}}{\tau {\rm \tanh}(|{\cal{Q}}_{l}B|\tau)}\,.
\end{equation}
In Fig.~\ref{Fig9}, we show numerical results for the average dynamical
quark mass in the chiral limit as a function of $eB$ (dotted line).
In the inset of Fig.~\ref{Fig9}, we plot again the average dynamical
quark mass in the chiral limit as a function of $eB$, but this time for
$eB$ up to $eB=10\,\text{GeV}^{2}$. We note that for large values of $eB$,
the dependence of $M$ on $eB$, is given by $\sqrt{eB}$; see the continuous
line in the inset of Fig.~\ref{Fig9}.

In Fig.~\ref{Fig10}, we present the evolution of dynamical average mass $M$
as a function of the number of flavors $N_f$ for various values of the
magnetic field $eB$. In this plot, we see that the value of $eB$, whose
magnetic catalysis enhances DCSB, competes against $N_f$ to generate dynamical
mass.
On the other hand, Fig.~\ref{Fig11} shows the behavior of the confinement
scale as a function of $N_f$.
From both figures, we can see that increasing $eB$ tends to increase the
$N_{f}^{\text{cr}}$ needed for chiral symmetry restoration, contrary
to the behavior of the critical number of fermions $N_{f}^{\text{cr}}$
as a function of temperature where it gets reduced as $T$ is increased~(see Fig.~\ref{Fig8}).

\begin{figure}[t!]
\begin{center}
\includegraphics[width=0.48\textwidth]{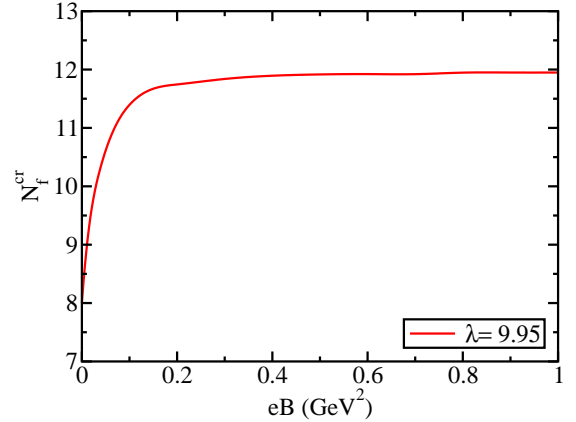}
\caption{ The phase diagram  of chiral symmetry and confinement for
$N_{f}^{\text{cr}}$ vs $eB$. $N_{f}^{\text{cr}}$ is obtained from the magnetic
gradient of the chiral condensate $\partial_{eB} \left\langle \bar{\psi} \psi
\right\rangle^{1/3} (N_f,eB)$ and of the confining scale $\partial_{eB}
\widetilde{\tau}_{\text{IR}}^{-1}(N_f,eB)$.}
\label{Fig12}
\end{center}
\end{figure}

That effect is clearer in Fig.~\ref{Fig12}, where we show the evolution of
critical $N_{f}^{\text{cr}}$ with respect to $eB$. These critical values are
obtained by locating the position of the divergences of their magnetic
gradients $\partial_{eB} M$ and
$\partial_{eB}\widetilde{\tau}_{\text{IR}}^{-1}$.
%
\section{Phase diagram at finite $T$ and $B$}
\begin{figure}[t!]
\begin{center}
\includegraphics[width=0.48\textwidth]{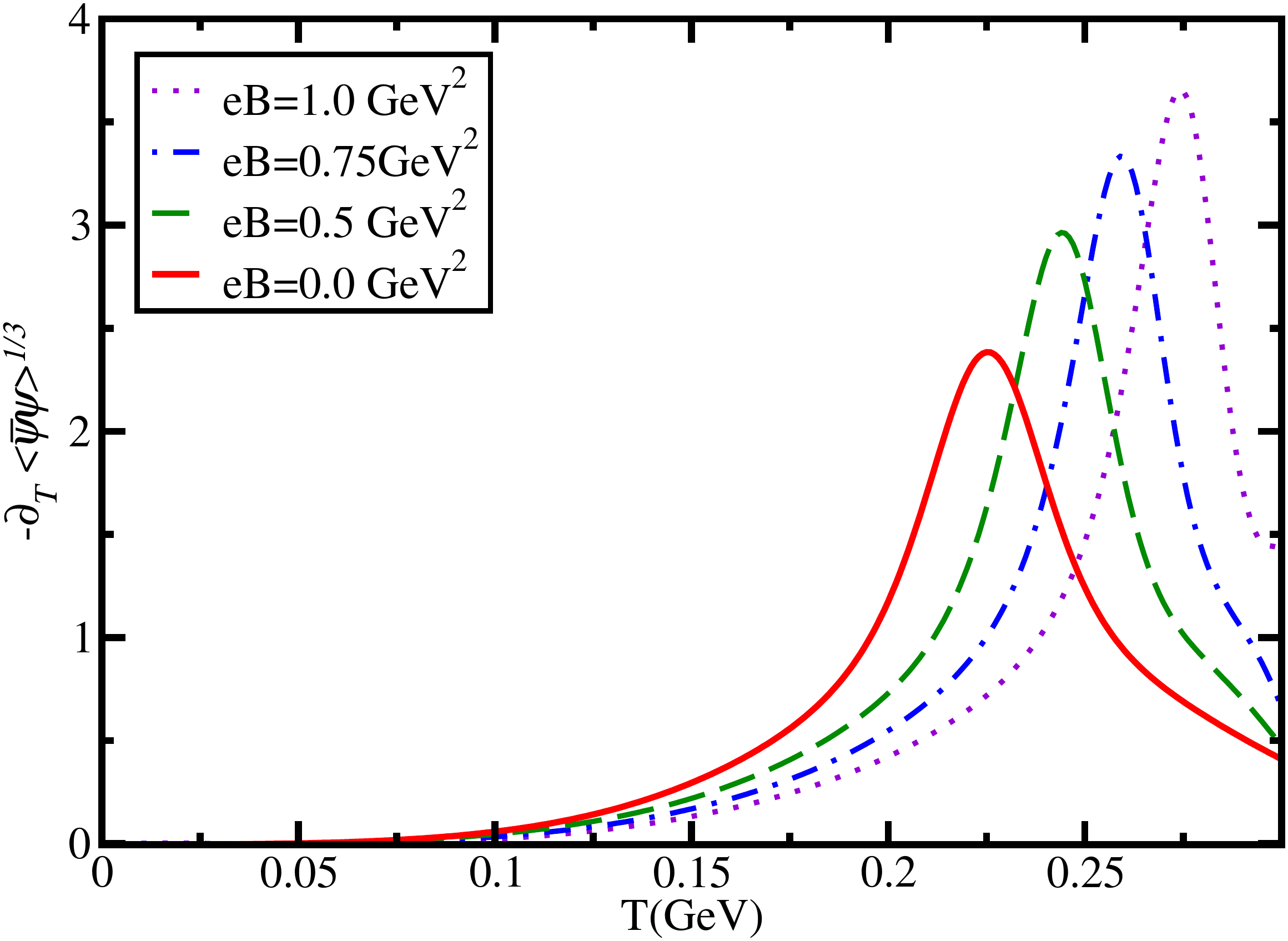}
\caption{The thermal gradient of the chiral condensate  $\partial_T \left\langle \bar{\psi} \psi
 \right\rangle^{1/3} (N_f=2,eB,T)$ has been plotted as a function of $T$ for various values of $eB$ using the magnetic field independent coupling.
It displays the usual feature of magnetic catalysis. For higher $T$, chiral symmetry breaking is triggered at larger values of $eB$.}
\label{Fig13}
\end{center}
\end{figure}

\begin{figure}[t!]
\begin{center}
\includegraphics[width=0.48\textwidth]{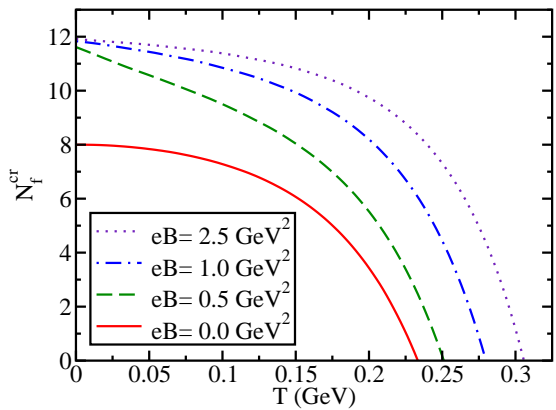}
\caption{The phase diagram  of chiral symmetry and confinement in the
$N_{f}^{\text{cr}}$ vs $T_{c}$ for various values of $eB$.
The $N_{f}^{\text{cr}}$ is obtained from the
thermal gradient of the chiral condensate  $\partial_T \left\langle \bar{\psi} \psi
 \right\rangle^{1/3} (N_f,eB,T)$ and of the confining scale
$\partial_T \widetilde{\tau}_{\text{IR}}^{-1}(N_f,eB,T)$.}
\label{Fig14}
\end{center}
\end{figure}

\begin{figure}[t!]
\begin{center}
\includegraphics[width=0.48\textwidth]{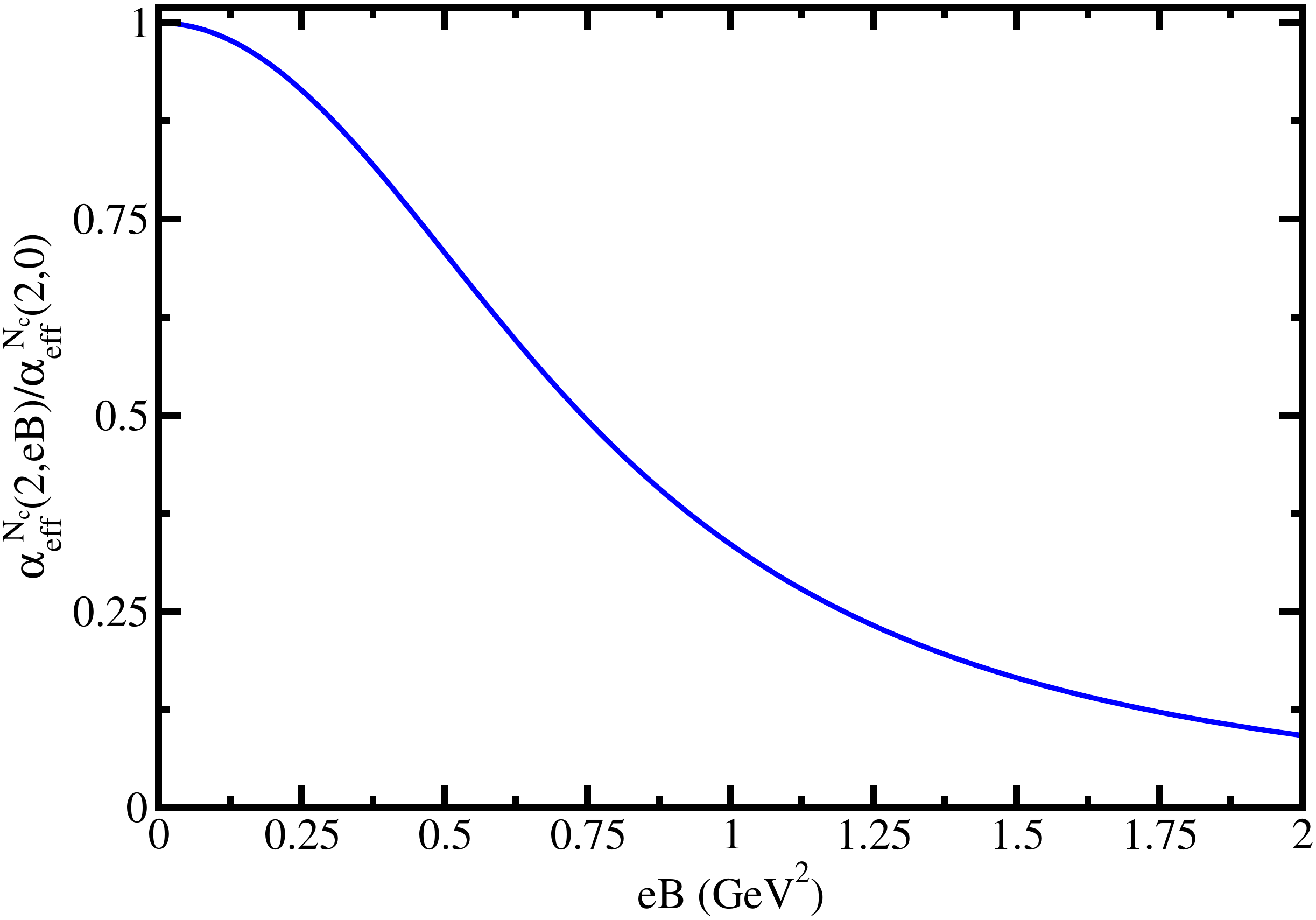}
\caption{Magnetic field dependent coupling of Eq.~(\ref{fceB}).}
\label{Fig15}
\end{center}
\end{figure}
At finite temperature $T$ and in a magnetic field $eB$, the gap equation beyond the chiral limit reads:
\begin{eqnarray}
M_{l}&=& m + \frac{16 \alpha_{\rm eff}(N_f)}{3} M_{l} T
\sum^{\infty}_{n=-\infty}\int\frac{d q_{3}}{(2\pi)}\frac{d^{2}
q_{\perp}}{(2\pi)^{2}}\nonumber\\ &\times&
\int_{\tau^{2}_{\text{UV}}}^{\widetilde\tau^{2}_{\text{IR}}}
d\tau {\rm e}^{-\tau(\omega^{2}_{n}+q^{2}_{3}+q^{2}_{\perp}
\frac{{\rm tan}(|{\cal Q}_{l}B\tau|)}{|{\cal Q}_{l}B\tau|}+M^{2}_{l})}
\label{MagT3},
\end{eqnarray}
for quark flavor $l=u,\,d$. We work for the isospin symmetric case, taking $m_u=m_d=m=7$MeV. After performing the sum over Matsubara frequencies, integrating over $q_3$ and $q_\perp$, the gap equation
of the average mass, Eq.~(\ref{eqn:mave}), at finite temperature and magnetic field is:
\begin{eqnarray}
M &=& m + \frac{2\alpha_{\rm eff}(N_f)MT}{3\pi^{3/2}}\frac{1}{2}
\sum_{l=u,d}
|{\cal Q}_{l}B| \nonumber \\
&\times& \int^{{\widetilde{\tau}}^2_{\text{IR}}}_{\tau^2_{\text{UV}}}d\tau
\frac{{\rm e}^{-M^2\tau}{\Theta_{2}(0,{\rm e}^{-(2\pi T)^{2}\tau})}}
{\tau^{1/2}\ {\rm \tanh}(|{\cal Q}_{l}B|\tau)},
\label{eqn:gapTBNf}
\end{eqnarray}
where
\begin{equation}
\widetilde{\tau}_{\text{IR}}=\tau_{\text{IR}} \frac{M(0,0,2)}{M(T,eB,N_f)}.
\end{equation}

The thermal gradient of the dynamical mass whose maximum points to the onslaught of DCSB has been plotted as a function of temperature for varying magnetic field
in Fig~{\ref{Fig13}}. We observe a typical pattern of magnetic catalysis. Increasing temperature requires larger magnetic field to catalyze DCSB.

In Fig.~\ref{Fig14}, we plot
$N_f^{cr}$ as a function of temperature for different values of the
magnetic field. We see the expected interplay between the magnetic field and the temperature: the greater the value of magnetic field, the higher is the temperature required to restore chiral symmetry. For any given value of the magnetic field strength, the critical value of the number of massless quark flavors decreases with the (critical)
temperature just as in the absence of the external magnetic field.
This pattern is again along the lines of the
common wisdom of magnetic catalysis. The only difference now is that the required $N_{f}^{\text{cr}}$
is somewhat larger for higher values of the magnetic field.

It is well known that the effect of the magnetic field must be taken into
account in the functional dependence of the interaction strength. If
$\alpha_{s}$ decreases with increasing $eB$ in a certain range of values of
the magnetic field, it suppresses the formation of chiral quark condensate.
This is what produces IMC.
In order to incorporate the magnetic field dependence in the coupling
constant and study its effect on $N_{f}^{\text{cr}}$, we follow~\cite{Ahmad:2016iez,Ferreira:2014kpa} and adopt the following
Pad\'e approximant for the
$eB$-dependent interaction strength
\begin{equation}
\alpha_{\rm eff}^{N_c}(N_{f},x)= \alpha_{\rm eff}^{N_c}(N_{f},0)
\left(\frac{1+ax^{2}+bx^{3}}{1+cx^{2}+dx^{4}}\right) \;,
\label{fceB}
\end{equation}
where the $N_{f}$ dependence of the coupling,
$\alpha_{\rm eff}^{N_c}(N_{f},0)$, is given by Eq.~(\ref{fc}),
$x=eB/\Lambda_{\text{QCD}}^2$, with $\Lambda_{\text{QCD}}=300$ MeV. The
parameters $a,b, c$ and $d$ were obtained~\cite{Ferreira:2014kpa} through reproducing
the critical transition temperature for chiral transition for different
values of the magnetic field strength, obtained by lattice QCD~\cite{Bali:2012zg}. The modified coupling is shown in
Fig.~\ref{Fig15}.
We have refrained from avoiding any additional $N_f$ dependence in this modified coupling.
We have no theoretical or phenomenological indications about the existence, let alone the the nature of such correlations between
$N_f$ and $eB$-dependence.

\begin{figure}[t!]
\begin{center}
\includegraphics[width=0.48\textwidth]{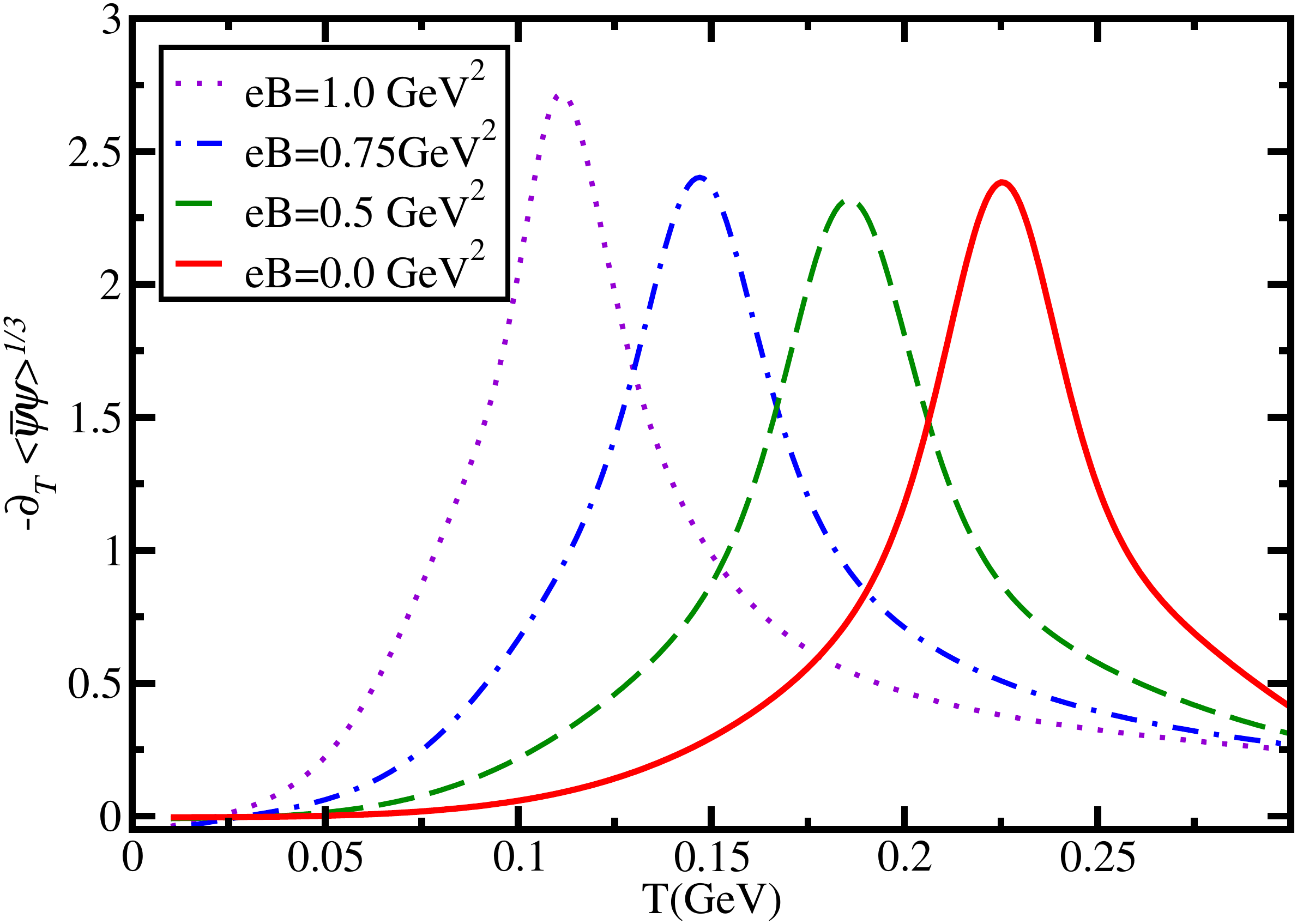}
\caption{The thermal gradient of the chiral condensate  $\partial_T \left\langle \bar{\psi} \psi
 \right\rangle^{1/3} (N_f=2,eB,T)$ has been plotted as a function of $T$ for various values of $eB$ using the magnetic field dependent coupling
of Eq.~(\ref{fceB}).
It now exhibits IMC. For higher $T$, chiral symmetry breaking occurs at lower values of $eB$.}
\label{Fig16}
\end{center}
\end{figure}

In Fig.~\ref{Fig16}, we redo the  plot
of thermal gradient of the dynamical mass against  temperature. We now observe a diametrically opposed behavior. Increasing temperature requires lower magnetic field for the onslaught of DCSB. This is
the well known IMC.

\begin{figure}[t!]
\begin{center}
\includegraphics[width=0.48\textwidth]{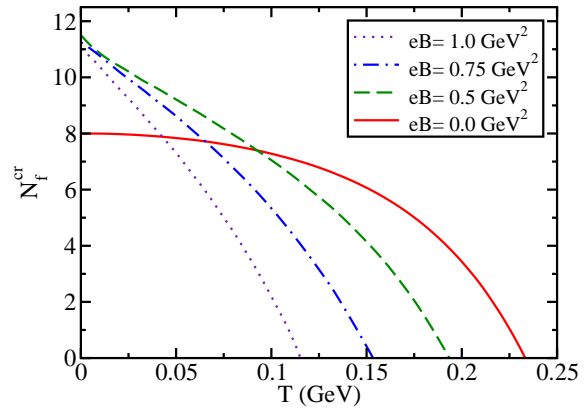}
\caption{The phase diagram  of chiral symmetry and confinement for
$N_{f}^{\text{cr}}$ vs $T_{c}$ at finite $eB$ obtained with an $eB$-dependent
coupling of Eq.~(\ref{fceB}).
The  $N_{f}^{\text{cr}}$ is
obtained from the thermal gradient of the chiral condensate
$\partial_T \left\langle \bar{\psi} \psi  \right\rangle^{1/3} (N_f,eB,T)$ and
of the confining scale
$\partial_T \widetilde{\tau}_{\text{IR}}^{-1}(N_f,eB,T)$}
\label{Fig17}
\end{center}
\end{figure}

In Fig.~\ref{Fig17}, we plot the new  $N_{f}-T$ phase diagram for varying magnetic field. As can be readily inferred from this figure,
the behaviour of the curves $N_f^{cr}$, as a function of $T$, is opposite
to that found in Fig.~\ref{Fig13}. This is another manifestation of the
IMC phenomenon predicted in this CI effective model. Lower values of $eB$ demand higher
$N_f^{cr}$ against the conventional wisdom of magnetic catalysis. Note that IMC
is not seen when temperatures are sufficiently low
as compared to the cross-over region.

\begin{figure}[t!]
\begin{center}
\includegraphics[width=0.48\textwidth]{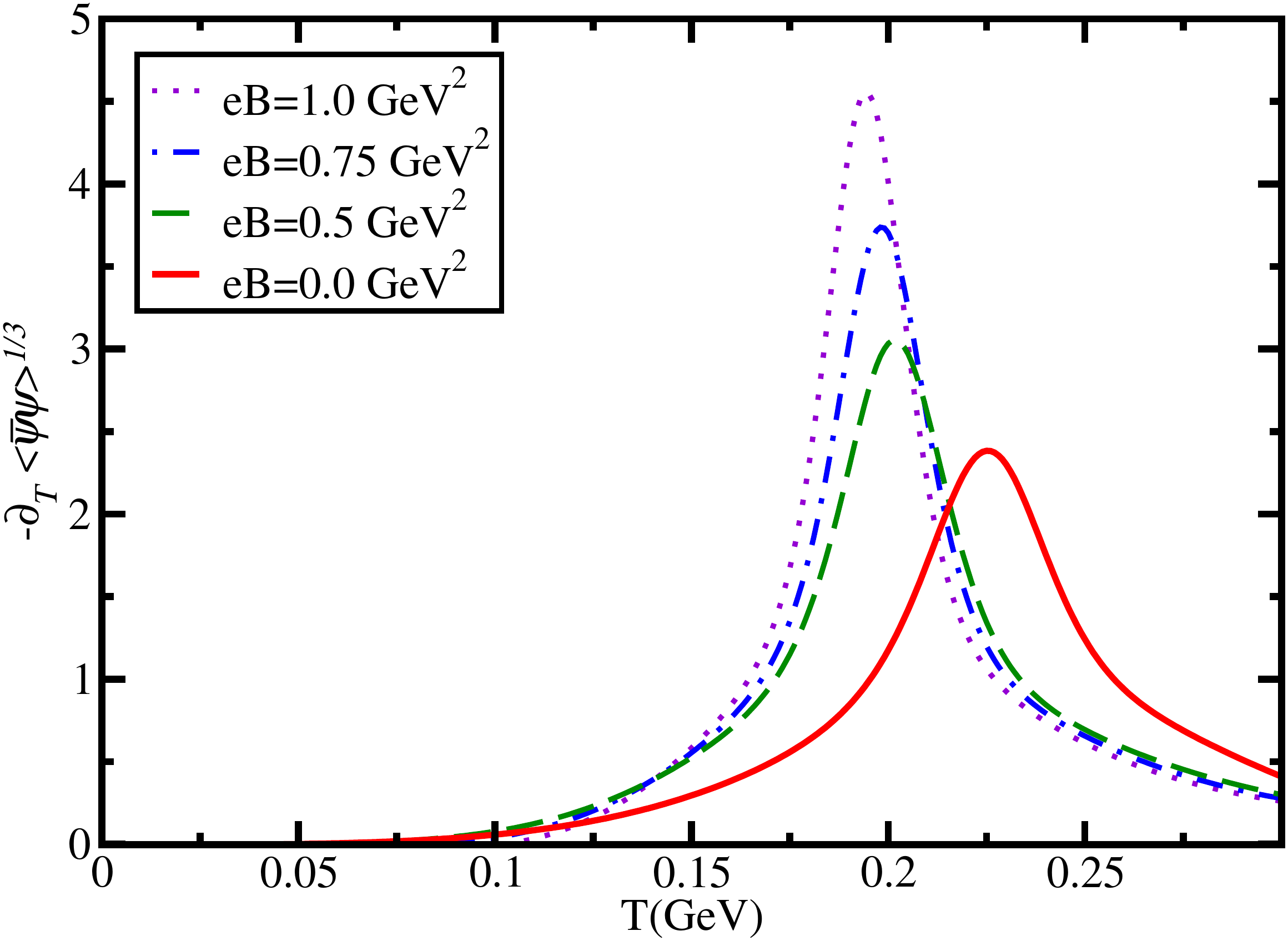}
\caption{The thermal gradient of the chiral condensate  $\partial_T \left\langle \bar{\psi} \psi
 \right\rangle^{1/3} (N_f=2,eB,T)$ has been plotted as a function of $T$ for various values of $eB$ using the magnetic field and temperature dependent coupling
of Eq.~(\ref{fceBT}).
It again exhibits IMC.}
\label{Fig18}
\end{center}
\end{figure}

\begin{figure}[t!]
\begin{center}
\includegraphics[width=0.48\textwidth]{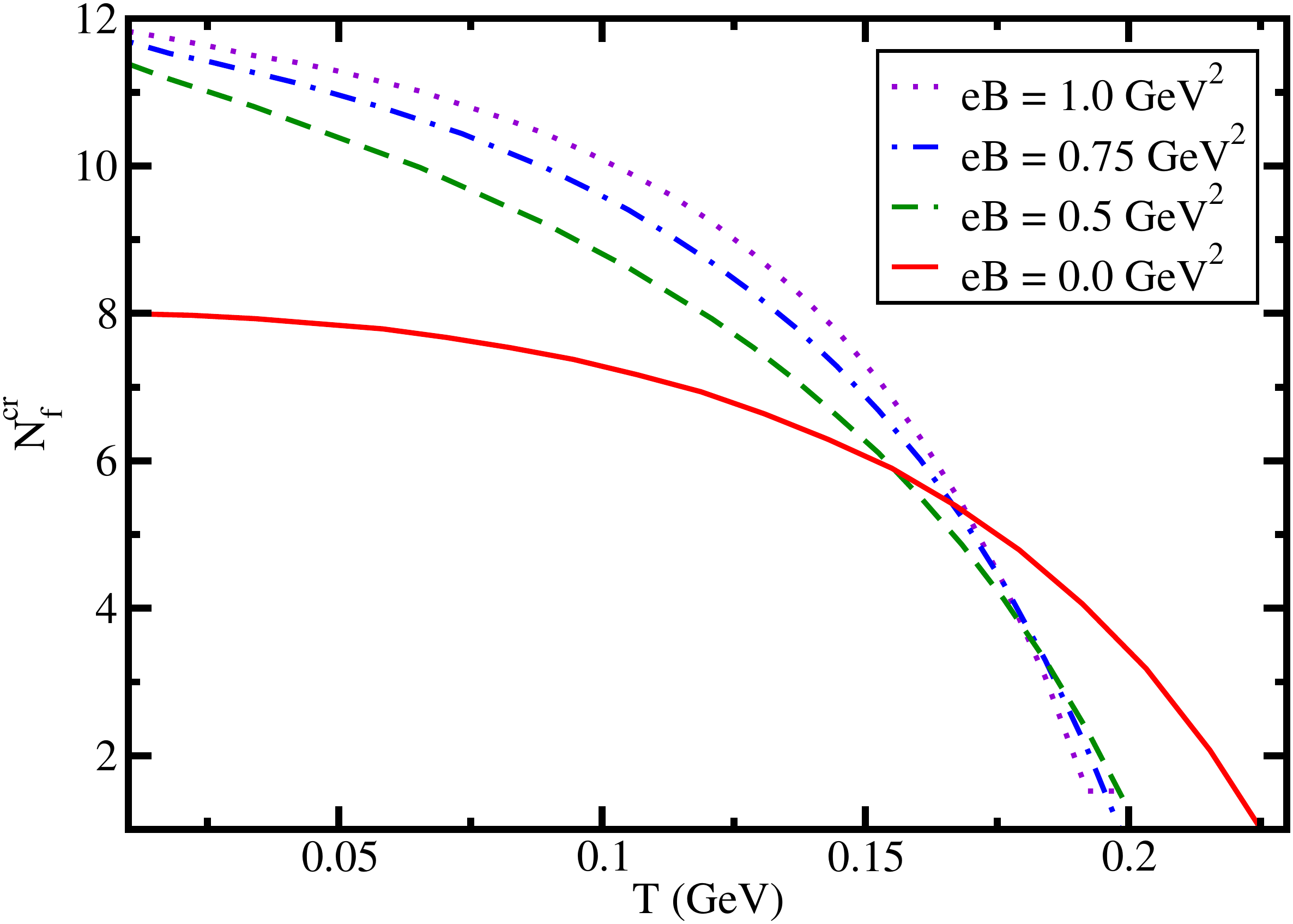}
\caption{The phase diagram  of chiral symmetry and confinement for
$N_{f}^{\text{cr}}$ vs $T_{c}$ at finite $eB$ obtained now with an $(eB,T)$-dependent
coupling of Eq.~(\ref{fceBT}).}
\label{Fig19}
\end{center}
\end{figure}

We can certainly do better than Eq.~(\ref{fceB}). The running coupling must also be a function of temperature.
This refined approach was adopted in~\cite{Farias:2014eca,Farias:2016gmy}. We follow~\cite{Farias:2014eca}, adopt their
parameters and define
 \begin{equation}
\alpha_{\rm eff}^{N_c}(N_{f},x,y)= \alpha_{\rm eff}^{N_c}(N_{f},0,0)
\frac{1 - \gamma x y}{1 + \alpha \, {\rm ln} (1+\beta x)}  \;,
\label{fceBT}
 \end{equation}
where $y=T/\Lambda_{\rm QCD}$. The parameters $\alpha$ and $\beta$ are fixed to obtain a reasonable description
of the lattice average of up and down quark condensates at $T=0$. $\gamma$ is obtained from a similar fit at the highest
temperatures. We now employ this coupling to show the corresponding plots in Fig.~\ref{Fig18} and Fig.~\ref{Fig19}.
Note that the IMC persists but in a narrower window near the cross-over temperature.

\section{Discussion and Conclusions}

We incorporate flavor dependence in our CI model to mimic
the latest SDE and lattice results. These results show that chiral symmetry
is restored and de-confinement is triggered above $N_f \approx 7 - 10$,~\cite{Appelquist:1996dq,Appelquist:1999hr,Bashir:2013zha,Hopfer:2014zna,Doff:2016jzk,Binosi:2016xxu}. This
observation is in accordance with the expectation that increasing light quark
flavors screen the interaction in contrast with the anti-screening effect of colors. For $N_f = 2$ (corresponding to $u$ and $d$ flavors), we find that the interaction is strong enough to break chiral symmetry only above $N_c \approx 2.2$. The nearest such integer is 3 which corresponds to the observed reality.

Additionally, the critical value $N^{\mathrm{cr}} _f$ decreases inversely with temperature. Temperature itself screens QCD interactions, thus ensuring even a small value of $N_f$ is sufficient to switch off DCSB.

Similarly, the presence of an external magnetic field also influences the value of $N^{\mathrm{cr}} _f$. $N^{\mathrm{cr}}_f$ grows with the increase of magnetic field as higher values of the later make it increasingly harder to pull the plug on
chiral symmetry breaking.

Last but not least, we explore the phase diagram for simultaneous variation of temperature and external magnetic field. We observe the usual magnetic catalysis for a strong enough magnetic field if the coupling strength is independent of the field. However, if we incorporate effective coupling of the model which decreases as a function of magnetic field in accordance with lattice studies, it triggers IMC in our model just as most modern
studies suggest. To the best of our knowledge, this work, for the first time, knits the quark flavor dependence with that of temperature and magnetic field in accordance with the current theoretical and phenomenological understanding of this field of study. We plan to investigate hadronic bound states within this model in future.  \\

\section{Acknowledgements}

We acknowledge financial support by CONACyT and CIC-UMSNH, Mexico, with research Grant Nos. CB-2014-242117
and 4.10, respectively.
We thank G. Krein and J. Rodr\'iguez-Quintero for their helpful comments on the
draft version of this article. A. Ahmad thanks A. Raya and P. Rioseco for their hospitality
during his stay at the IFM, UMSNH, Mexico.


\begin{thebibliography}{99}

\bibitem{Gross:1973id}
  D.~J.~Gross and F.~Wilczek,
  Phys.\ Rev.\ Lett.\  {\bf 30}, 1343 (1973).
  doi:10.1103/PhysRevLett.30.1343

\bibitem{Politzer:1973fx}
  H.~D.~Politzer,
  Phys.\ Rev.\ Lett.\  {\bf 30}, 1346 (1973).
  doi:10.1103/PhysRevLett.30.1346

\bibitem{Gross:1973ju}
  D.~J.~Gross and F.~Wilczek,
  Phys.\ Rev.\ D {\bf 8}, 3633 (1973).
  doi:10.1103/PhysRevD.8.3633

\bibitem{Hayakawa:2010yn}
  M.~Hayakawa, K.-I.~Ishikawa, Y.~Osaki, S.~Takeda, S.~Uno and N.~Yamada,
  Phys.\ Rev.\ D {\bf 83}, 074509 (2011)
  doi:10.1103/PhysRevD.83.074509
  [arXiv:1011.2577 [hep-lat]].

\bibitem{Cheng:2013eu}
  A.~Cheng, A.~Hasenfratz, G.~Petropoulos and D.~Schaich,
  JHEP {\bf 1307}, 061 (2013)
  doi:10.1007/JHEP07(2013)061
  [arXiv:1301.1355 [hep-lat]].

\bibitem{Appelquist:2014zsa}
  T.~Appelquist {\it et al.} [LSD Collaboration],
  Phys.\ Rev.\ D {\bf 90}, no. 11, 114502 (2014)
  doi:10.1103/PhysRevD.90.114502
  [arXiv:1405.4752 [hep-lat]].

\bibitem{Hasenfratz:2016dou}
  A.~Hasenfratz and D.~Schaich,
  JHEP {\bf 1802}, 132 (2018)
  doi:10.1007/JHEP02(2018)132
  [arXiv:1610.10004 [hep-lat]].

\bibitem{Appelquist:2018yqe}
  T.~Appelquist {\it et al.} [Lattice Strong Dynamics Collaboration],
  Phys.\ Rev.\ D {\bf 99}, no. 1, 014509 (2019)
  doi:10.1103/PhysRevD.99.014509
  [arXiv:1807.08411 [hep-lat]].


\bibitem{Appelquist:1996dq}
  T.~Appelquist, J.~Terning and L.~C.~R.~Wijewardhana,
  Phys.\ Rev.\ Lett.\  {\bf 77}, 1214 (1996)
  doi:10.1103/PhysRevLett.77.1214
  [hep-ph/9602385].

\bibitem{Appelquist:1999hr}
  T.~Appelquist, A.~G.~Cohen and M.~Schmaltz,
  Phys.\ Rev.\ D {\bf 60}, 045003 (1999)
  doi:10.1103/PhysRevD.60.045003
  [hep-th/9901109].

\bibitem{Bashir:2013zha}
  A.~Bashir, A.~Raya and J.~Rodriguez-Quintero,
  Phys.\ Rev.\ D {\bf 88}, 054003 (2013)
  doi:10.1103/PhysRevD.88.054003
  [arXiv:1302.5829 [hep-ph]].

\bibitem{Hopfer:2014zna}
  M.~Hopfer, C.~S.~Fischer and R.~Alkofer,
  JHEP {\bf 1411}, 035 (2014)
  doi:10.1007/JHEP11(2014)035
  [arXiv:1405.7031 [hep-ph]].

\bibitem{Doff:2016jzk}
  A.~Doff and A.~A.~Natale,
  Phys.\ Rev.\ D {\bf 94}, no. 7, 076005 (2016)
  doi:10.1103/PhysRevD.94.076005
  [arXiv:1610.02564 [hep-ph]].

\bibitem{Binosi:2016xxu}
  D.~Binosi, C.~D.~Roberts and J.~Rodriguez-Quintero,
  Phys.\ Rev.\ D {\bf 95}, no. 11, 114009 (2017)
  doi:10.1103/PhysRevD.95.114009
  [arXiv:1611.03523 [nucl-th]].


\bibitem{Bernard:2004je}
  C.~Bernard {\it et al.} [MILC Collaboration],
  Phys.\ Rev.\ D {\bf 71}, 034504 (2005)
  doi:10.1103/PhysRevD.71.034504
  [hep-lat/0405029].

\bibitem{Cheng:2006qk}
  M.~Cheng {\it et al.},
  Phys.\ Rev.\ D {\bf 74}, 054507 (2006)
  doi:10.1103/PhysRevD.74.054507
  [hep-lat/0608013].

\bibitem{Aoki:2009sc}
  Y.~Aoki, S.~Borsanyi, S.~Durr, Z.~Fodor, S.~D.~Katz, S.~Krieg and K.~K.~Szabo,
  JHEP {\bf 0906}, 088 (2009)
  doi:10.1088/1126-6708/2009/06/088
  [arXiv:0903.4155 [hep-lat]].

\bibitem{Borsanyi:2010bp}
  S.~Borsanyi {\it et al.} [Wuppertal-Budapest Collaboration],
  JHEP {\bf 1009}, 073 (2010)
  doi:10.1007/JHEP09(2010)073
  [arXiv:1005.3508 [hep-lat]].

\bibitem{Bazavov:2011nk}
  A.~Bazavov {\it et al.},
  Phys.\ Rev.\ D {\bf 85}, 054503 (2012)
  doi:10.1103/PhysRevD.85.054503
  [arXiv:1111.1710 [hep-lat]].

\bibitem{Levkova:2012jd}
  L.~Levkova,
  PoS LATTICE {\bf 2011}, 011 (2011)
  doi:10.22323/1.139.0011
  [arXiv:1201.1516 [hep-lat]].

\bibitem{Bhattacharya:2014ara}
  T.~Bhattacharya {\it et al.},
  Phys.\ Rev.\ Lett.\  {\bf 113}, no. 8, 082001 (2014)
  doi:10.1103/PhysRevLett.113.082001
  [arXiv:1402.5175 [hep-lat]].

\bibitem{deForcrand:2014tha}
  P.~de Forcrand, J.~Langelage, O.~Philipsen and W.~Unger,
  Phys.\ Rev.\ Lett.\  {\bf 113}, no. 15, 152002 (2014)
  doi:10.1103/PhysRevLett.113.152002
  [arXiv:1406.4397 [hep-lat]].

\bibitem{Bazavov:2017xul}
  A.~Bazavov, H.-T.~Ding, P.~Hegde, F.~Karsch, E.~Laermann, S.~Mukherjee, P.~Petreczky and C.~Schmidt,
  Phys.\ Rev.\ D {\bf 95}, no. 7, 074505 (2017)
  doi:10.1103/PhysRevD.95.074505
  [arXiv:1701.03548 [hep-lat]].



\bibitem{Qin:2010nq}
  S.~x.~Qin, L.~Chang, H.~Chen, Y.~x.~Liu and C.~D.~Roberts,
  Phys.\ Rev.\ Lett.\  {\bf 106}, 172301 (2011)
  doi:10.1103/PhysRevLett.106.172301
  [arXiv:1011.2876 [nucl-th]].

\bibitem{Fischer:2011mz}
  C.~S.~Fischer, J.~Luecker and J.~A.~Mueller,
  Phys.\ Lett.\ B {\bf 702}, 438 (2011)
  doi:10.1016/j.physletb.2011.07.039
  [arXiv:1104.1564 [hep-ph]].

\bibitem{Ayala:2011vs}
  A.~Ayala, A.~Bashir, C.~A.~Dominguez, E.~Gutierrez, M.~Loewe and A.~Raya,
  Phys.\ Rev.\ D {\bf 84}, 056004 (2011)
  doi:10.1103/PhysRevD.84.056004
  [arXiv:1106.5155 [hep-ph]].

\bibitem{Gutierrez:2013sta}
  E.~Gutierrez, A.~Ahmad, A.~Ayala, A.~Bashir and A.~Raya,
  J.\ Phys.\ G {\bf 41}, 075002 (2014)
  doi:10.1088/0954-3899/41/7/075002
  [arXiv:1304.8065 [hep-ph]].

\bibitem{Gao:2015kea}
  F.~Gao, J.~Chen, Y.~X.~Liu, S.~X.~Qin, C.~D.~Roberts and S.~M.~Schmidt,
  Phys.\ Rev.\ D {\bf 93}, no. 9, 094019 (2016)
  doi:10.1103/PhysRevD.93.094019
  [arXiv:1507.00875 [nucl-th]].

\bibitem{Eichmann:2015kfa}
  G.~Eichmann, C.~S.~Fischer and C.~A.~Welzbacher,
  Phys.\ Rev.\ D {\bf 93}, no. 3, 034013 (2016)
  doi:10.1103/PhysRevD.93.034013
  [arXiv:1509.02082 [hep-ph]].

\bibitem{Gao:2016qkh}
  F.~Gao and Y.~x.~Liu,
  Phys.\ Rev.\ D {\bf 94}, no. 7, 076009 (2016)
  doi:10.1103/PhysRevD.94.076009
  [arXiv:1607.01675 [hep-ph]].

\bibitem{Fischer:2018sdj}
  C.~S.~Fischer,
  Prog.\ Part.\ Nucl.\ Phys.\  {\bf 105}, 1 (2019)
  doi:10.1016/j.ppnp.2019.01.002
  [arXiv:1810.12938 [hep-ph]].

\bibitem{Shi:2020uyb}
C.~Shi, X.~T.~He, W.~B.~Jia, Q.~W.~Wang, S.~S.~Xu and H.~S.~Zong,
JHEP \textbf{06}, 122 (2020)
doi:10.1007/JHEP06(2020)122
[arXiv:2004.09918 [hep-ph]].


\bibitem{Gusynin:1994xp}
  V.~P.~Gusynin, V.~A.~Miransky and I.~A.~Shovkovy,
  Phys.\ Lett.\ B {\bf 349}, 477 (1995)
  doi:10.1016/0370-2693(95)00232-A
  [hep-ph/9412257].



\bibitem{Lee:1997zj}
  D.~S.~Lee, C.~N.~Leung and Y.~J.~Ng,
  Phys.\ Rev.\ D {\bf 55}, 6504 (1997)
  doi:10.1103/PhysRevD.55.6504
  [hep-th/9701172].



\bibitem{Hong:1997uw}
  D.~K.~Hong,
  Phys.\ Rev.\ D {\bf 57}, 3759 (1998)
  doi:10.1103/PhysRevD.57.3759
  [hep-ph/9707432].



\bibitem{Ferrer:2000ed}
  E.~J.~Ferrer and V.~de la Incera,
  Phys.\ Lett.\ B {\bf 481}, 287 (2000)
  doi:10.1016/S0370-2693(00)00482-2
  [hep-ph/0004113].



\bibitem{Ayala:2006sv}
  A.~Ayala, A.~Bashir, A.~Raya and E.~Rojas,
  Phys.\ Rev.\ D {\bf 73}, 105009 (2006)
  doi:10.1103/PhysRevD.73.105009
  [hep-ph/0602209].


\bibitem{Rojas:2008sg}
  E.~Rojas, A.~Ayala, A.~Bashir and A.~Raya,
  Phys.\ Rev.\ D {\bf 77}, 093004 (2008)
  doi:10.1103/PhysRevD.77.093004
  [arXiv:0803.4173 [hep-ph]].


\bibitem{Ayala:2010fm}
  A.~Ayala, A.~Bashir, E.~Gutierrez, A.~Raya and A.~Sanchez,
  Phys.\ Rev.\ D {\bf 82}, 056011 (2010)
  doi:10.1103/PhysRevD.82.056011
  [arXiv:1007.4249 [hep-ph]].


\bibitem{Kabat:2002er}
  D.~N.~Kabat, K.~M.~Lee and E.~J.~Weinberg,
  Phys.\ Rev.\ D {\bf 66}, 014004 (2002)
  doi:10.1103/PhysRevD.66.014004
  [hep-ph/0204120].



\bibitem{Miransky:2002rp}
  V.~A.~Miransky and I.~A.~Shovkovy,
  Phys.\ Rev.\ D {\bf 66}, 045006 (2002)
  doi:10.1103/PhysRevD.66.045006
  [hep-ph/0205348].

\bibitem{DElia:2010abb}
  M.~D'Elia, S.~Mukherjee and F.~Sanfilippo,
  Phys.\ Rev.\ D {\bf 82}, 051501 (2010)
  doi:10.1103/PhysRevD.82.051501
  [arXiv:1005.5365 [hep-lat]].

\bibitem{Bali:2011qj}
  G.~S.~Bali, F.~Bruckmann, G.~Endrodi, Z.~Fodor, S.~D.~Katz, S.~Krieg, A.~Schafer and K.~K.~Szabo,
  JHEP {\bf 1202}, 044 (2012)
  doi:10.1007/JHEP02(2012)044
  [arXiv:1111.4956 [hep-lat]].

\bibitem{Bali:2012zg}
  G.~S.~Bali, F.~Bruckmann, G.~Endrodi, Z.~Fodor, S.~D.~Katz and A.~Schafer,
  Phys.\ Rev.\ D {\bf 86}, 071502 (2012)
  doi:10.1103/PhysRevD.86.071502
  [arXiv:1206.4205 [hep-lat]].

\bibitem{Bali:2013esa}
  G.~S.~Bali, F.~Bruckmann, G.~Endrodi, F.~Gruber and A.~Schaefer,
  JHEP {\bf 1304}, 130 (2013)
  doi:10.1007/JHEP04(2013)130
  [arXiv:1303.1328 [hep-lat]].

\bibitem{Bornyakov:2013eya}
  V.~G.~Bornyakov, P.~V.~Buividovich, N.~Cundy, O.~A.~Kochetkov and A.~Schäfer,
  Phys.\ Rev.\ D {\bf 90}, no. 3, 034501 (2014)
  doi:10.1103/PhysRevD.90.034501
  [arXiv:1312.5628 [hep-lat]].

\bibitem{Pagura:2016pwr}
V.~P.~Pagura, D.~Gomez Dumm, S.~Noguera and N.~N.~Scoccola,
Phys. Rev. D \textbf{95} (2017) no.3, 034013
doi:10.1103/PhysRevD.95.034013
[arXiv:1609.02025 [hep-ph]].


\bibitem{Fukushima:2012kc}
  K.~Fukushima and Y.~Hidaka,
  Phys.\ Rev.\ Lett.\  {\bf 110}, no. 3, 031601 (2013)
  doi:10.1103/PhysRevLett.110.031601
  [arXiv:1209.1319 [hep-ph]].

\bibitem{Mueller:2015fka}
  N.~Mueller and J.~M.~Pawlowski,
  Phys.\ Rev.\ D {\bf 91}, no. 11, 116010 (2015)
  doi:10.1103/PhysRevD.91.116010
  [arXiv:1502.08011 [hep-ph]].

\bibitem{GutierrezGuerrero:2010md}
  L.~X.~Gutierrez-Guerrero, A.~Bashir, I.~C.~Cloet and C.~D.~Roberts,
  Phys.\ Rev.\ C {\bf 81}, 065202 (2010)
  doi:10.1103/PhysRevC.81.065202
  [arXiv:1002.1968 [nucl-th]].

\bibitem{Roberts:2010rn}
  H.~L.~L.~Roberts, C.~D.~Roberts, A.~Bashir, L.~X.~Gutierrez-Guerrero and P.~C.~Tandy,
  Phys.\ Rev.\ C {\bf 82}, 065202 (2010)
  doi:10.1103/PhysRevC.82.065202
  [arXiv:1009.0067 [nucl-th]].

\bibitem{Roberts:2011wy}
  H.~L.~L.~Roberts, A.~Bashir, L.~X.~Gutierrez-Guerrero, C.~D.~Roberts and D.~J.~Wilson,
  Phys.\ Rev.\ C {\bf 83}, 065206 (2011)
  doi:10.1103/PhysRevC.83.065206
  [arXiv:1102.4376 [nucl-th]].

\bibitem{Chen:2012qr}
  C.~Chen, L.~Chang, C.~D.~Roberts, S.~Wan and D.~J.~Wilson,
  Few Body Syst.\  {\bf 53}, 293 (2012)
  doi:10.1007/s00601-012-0466-3
  [arXiv:1204.2553 [nucl-th]].

\bibitem{Roberts:2011cf}
  H.~L.~L.~Roberts, L.~Chang, I.~C.~Cloet and C.~D.~Roberts,
  Few Body Syst.\  {\bf 51}, 1 (2011)
  doi:10.1007/s00601-011-0225-x
  [arXiv:1101.4244 [nucl-th]].

\bibitem{Bedolla:2015mpa}
  M.~A.~Bedolla, J.~J.~Cobos-Mart\'inez and A.~Bashir,
  Phys.\ Rev.\ D {\bf 92}, no. 5, 054031 (2015)
  doi:10.1103/PhysRevD.92.054031
  [arXiv:1601.05639 [hep-ph]].

\bibitem{Bedolla:2016yxq}
  M.~A.~Bedolla, K.~Raya, J.~J.~Cobos-Mart\'inez and A.~Bashir,
  Phys.\ Rev.\ D {\bf 93}, no. 9, 094025 (2016)
  doi:10.1103/PhysRevD.93.094025
  [arXiv:1606.03760 [hep-ph]].

\bibitem{Raya:2017ggu}
  K.~Raya, M.~A.~Bedolla, J.~J.~Cobos-Mart\'inez and A.~Bashir,
  Few Body Syst.\  {\bf 59}, no. 6, 133 (2018)
  doi:10.1007/s00601-018-1455-y
  [arXiv:1711.00383 [nucl-th]].

\bibitem{Gutierrez-Guerrero:2019uwa}
L.~X.~Guti\'errez-Guerrero, A.~Bashir, M.~A.~Bedolla and E.~Santopinto,
Phys. Rev. D \textbf{100}, no.11, 114032 (2019)
doi:10.1103/PhysRevD.100.114032
[arXiv:1911.09213 [nucl-th]].

\bibitem{Bowman:2004jm}
  P.~O.~Bowman, U.~M.~Heller, D.~B.~Leinweber, M.~B.~Parappilly and A.~G.~Williams,
  Phys.\ Rev.\ D {\bf 70}, 034509 (2004)
  doi:10.1103/PhysRevD.70.034509
  [hep-lat/0402032].

\bibitem{Dudal:2008sp}
  D.~Dudal, J.~A.~Gracey, S.~P.~Sorella, N.~Vandersickel and H.~Verschelde,
  Phys.\ Rev.\ D {\bf 78} (2008) 065047
  doi:10.1103/PhysRevD.78.065047
  [arXiv:0806.4348 [hep-th]].

\bibitem{Huber:2010cq}
  M.~Q.~Huber, R.~Alkofer and S.~P.~Sorella,
  AIP Conf.\ Proc.\  {\bf 1343}, 158 (2011)
  doi:10.1063/1.3574962
  [arXiv:1010.4802 [hep-th]].

\bibitem{Boucaud:2011ug}
  P.~Boucaud, J.~P.~Leroy, A.~L.~Yaouanc, J.~Micheli, O.~Pene and J.~Rodriguez-Quintero,
  Few Body Syst.\  {\bf 53}, 387 (2012)
  doi:10.1007/s00601-011-0301-2
  [arXiv:1109.1936 [hep-ph]].

\bibitem{Ayala:2012pb}
  A.~Ayala, A.~Bashir, D.~Binosi, M.~Cristoforetti and J.~Rodriguez-Quintero,
  Phys.\ Rev.\ D {\bf 86}, 074512 (2012)
  doi:10.1103/PhysRevD.86.074512
  [arXiv:1208.0795 [hep-ph]].

\bibitem{Ahmad:2016iez}
  A.~Ahmad and A.~Raya,
  J.\ Phys.\ G {\bf 43}, no. 6, 065002 (2016)
  doi:10.1088/0954-3899/43/6/065002
  [arXiv:1602.06448 [hep-ph]].

\bibitem{Gao:2017uox}
  F.~Gao, S.~X.~Qin, C.~D.~Roberts and J.~Rodriguez-Quintero,
  Phys.\ Rev.\ D {\bf 97} (2018) no.3,  034010
  doi:10.1103/PhysRevD.97.034010
  [arXiv:1706.04681 [hep-ph]].

\bibitem{Ebert:1996vx}
  D.~Ebert, T.~Feldmann and H.~Reinhardt,
  Phys.\ Lett.\ B {\bf 388}, 154 (1996)
  doi:10.1016/0370-2693(96)01158-6
  [hep-ph/9608223].

\bibitem{Roberts:2007ji}
  C.~D.~Roberts,
  Prog.\ Part.\ Nucl.\ Phys.\  {\bf 61}, 50 (2008)
  doi:10.1016/j.ppnp.2007.12.034
  [arXiv:0712.0633 [nucl-th]].


\bibitem{Marquez:2015bca}
  F.~Marquez, A.~Ahmad, M.~Buballa and A.~Raya,
  Phys.\ Lett.\ B {\bf 747}, 529 (2015)
  doi:10.1016/j.physletb.2015.06.031
  [arXiv:1504.06730 [nucl-th]].

\bibitem{Cui:2014hya}
  Z.~F.~Cui, Y.~L.~Du and H.~S.~Zong,
  Int.\ J.\ Mod.\ Phys.\ Conf.\ Ser.\  {\bf 29}, 1460232 (2014).
  doi:10.1142/S2010194514602324


\bibitem{Wang:2013wk}
  K.~l.~Wang, Y.~x.~Liu, L.~Chang, C.~D.~Roberts and S.~M.~Schmidt,
  Phys.\ Rev.\ D {\bf 87}, no. 7, 074038 (2013)
  doi:10.1103/PhysRevD.87.074038
  [arXiv:1301.6762 [nucl-th]].



\bibitem{Schwinger:1951nm}
  J.~S.~Schwinger,
  Phys.\ Rev.\  {\bf 82}, 664 (1951).
  doi:10.1103/PhysRev.82.664


\bibitem{Fock:1937dy}
  V.~Fock,
  Phys.\ Z.\ Sowjetunion {\bf 12}, 404 (1937).


\bibitem{Ferreira:2014kpa}
M.~Ferreira, P.~Costa, O.~Lourenço, T.~Frederico and C.~Providência,
Phys. Rev. D \textbf{89}, no.11, 116011 (2014)
doi:10.1103/PhysRevD.89.116011
[arXiv:1404.5577 [hep-ph]].

\bibitem{Farias:2014eca}
R.~L.~S.~Farias, K.~P.~Gomes, G.~I.~Krein and M.~B.~Pinto,
Phys. Rev. C \textbf{90}, no.2, 025203 (2014)
doi:10.1103/PhysRevC.90.025203
[arXiv:1404.3931 [hep-ph]].

\bibitem{Farias:2016gmy}
R.~L.~S.~Farias, V.~S.~Timoteo, S.~S.~Avancini, M.~B.~Pinto and G.~Krein,
Eur. Phys. J. A \textbf{53}, no.5, 101 (2017)
doi:10.1140/epja/i2017-12320-8
[arXiv:1603.03847 [hep-ph]].

\end{thebibliography}
\end{document}